\input harvmac
\let\includefigures=\iftrue
\let\useblackboard==\iftrue
\newfam\black

\includefigures
\message{If you do not have epsf.tex (to include figures),}
\message{change the option at the top of the tex file.}
\input epsf
\def\figin{\epsfcheck\figin}\def\figins{\epsfcheck\figins}
\def\epsfcheck{\ifx\epsfbox\UnDeFiNeD
\message{(NO epsf.tex, FIGURES WILL BE IGNORED)}
\gdef\figin##1{\vskip2in}\gdef\figins##1{\hskip.5in}
\else\message{(FIGURES WILL BE INCLUDED)}%
\gdef\figin##1{##1}\gdef\figins##1{##1}\fi}
\def\DefWarn#1{}
\def\figinsert{\goodbreak\midinsert}
\def\ifig#1#2#3{\DefWarn#1\xdef#1{fig.~\the\figno}
\writedef{#1\leftbracket fig.\noexpand~\the\figno}%
\figinsert\figin{\centerline{#3}}\medskip\centerline{\vbox{
\baselineskip12pt\advance\hsize by -1truein
\noindent\footnotefont{\bf Fig.~\the\figno:} #2}}
\endinsert\global\advance\figno by1}
\else
\def\ifig#1#2#3{\xdef#1{fig.~\the\figno}
\writedef{#1\leftbracket fig.\noexpand~\the\figno}%
\global\advance\figno by1} \fi

\def\id{{1 \kern-.28em {\rm l}}}
\def\N{{\cal N}}
\def\O{{\cal O}}

\def\K3{{\bf K3}}
\def\journal#1&#2(#3){\unskip, \sl #1\ \bf #2 \rm(19#3) }
\def\andjournal#1&#2(#3){\sl #1~\bf #2 \rm (19#3) }

\def\bar{\overline}
\def\hat{\widehat}
\def\ie{{\it i.e.}}

\def\tilde{\widetilde}

\def\frac#1#2{{#1\over#2}}

\def\half{\frac12}

\def\inbar{\,\vrule height1.5ex width.4pt depth0pt}
\def\IC{\relax\hbox{$\inbar\kern-.3em{\rm C}$}}
\def\IR{\relax{\rm I\kern-.18em R}}
\def\IP{\relax{\rm I\kern-.18em P}}

%
%

%
\catcode`\@=11
\def\slash#1{\mathord{\mathpalette\c@ncel{#1}}}
\overfullrule=0pt

\def\underrel#1\over#2{\mathrel{\mathop{\kern\z@#1}\limits_{#2}}}

\catcode`\@=12


%

\def\O{{\cal O}}

\def\N{{\cal N}}
\def\ie{{\it i.e.}}

\lref\IntriligatorMI{
  K.~A.~Intriligator and B.~Wecht,
  ``RG fixed points and flows in SQCD with adjoints,''
Nucl.\ Phys.\ B {\bf 677}, 223 (2004).
[hep-th/0309201].
}

\lref\VenezianoEC{
  G.~Veneziano,
  ``U(1) Without Instantons,''
Nucl.\ Phys.\ B {\bf 159}, 213 (1979)..
}

\lref\BanksNN{
  T.~Banks and A.~Zaks,
  ``On the Phase Structure of Vector-Like Gauge Theories with Massless Fermions,''
Nucl.\ Phys.\ B {\bf 196}, 189 (1982)..
}
\lref\PapadimitriouAP{
  I.~Papadimitriou and K.~Skenderis,
  ``AdS / CFT correspondence and geometry,''
IRMA Lect.\ Math.\ Theor.\ Phys.\  {\bf 8}, 73 (2005).
[hep-th/0404176].
}

\lref\CardyCWA{
  J.~L.~Cardy,
  ``Is There a c Theorem in Four-Dimensions?,''
Phys.\ Lett.\ B {\bf 215}, 749 (1988)..
}

\lref\HerzogED{
  C.~P.~Herzog and K.~W.~Huang,
  ``Stress Tensors from Trace Anomalies in Conformal Field Theories,''
Phys.\ Rev.\ D {\bf 87}, 081901 (2013).
[arXiv:1301.5002 [hep-th]].
}

\lref\KutasovNP{
  D.~Kutasov and A.~Schwimmer,
  ``On duality in supersymmetric Yang-Mills theory,''
Phys.\ Lett.\ B {\bf 354}, 315 (1995).
[hep-th/9505004].
}
\lref\KutasovSS{
  D.~Kutasov, A.~Schwimmer and N.~Seiberg,
  ``Chiral rings, singularity theory and electric - magnetic duality,''
Nucl.\ Phys.\ B {\bf 459}, 455 (1996).
[hep-th/9510222].
}

\lref\SeibergPQ{
  N.~Seiberg,
  ``Electric - magnetic duality in supersymmetric non-Abelian gauge theories,''
Nucl.\ Phys.\ B {\bf 435}, 129 (1995).
[hep-th/9411149].
}

\lref\StrasslerQS{
  M.~J.~Strassler,
  ``The Duality cascade,''
[hep-th/0505153].
}

\lref\StrasslerQG{
  M.~J.~Strassler,
  ``An Unorthodox introduction to supersymmetric gauge theory,''
[hep-th/0309149].
}

\lref\KutasovXU{
  D.~Kutasov and A.~Schwimmer,
  ``Lagrange multipliers and couplings in supersymmetric field theory,''
Nucl.\ Phys.\ B {\bf 702}, 369 (2004).
[hep-th/0409029].
}

\lref\IntriligatorXXA{
  K.~Intriligator and F.~Sannino,
  ``Supersymmetric asymptotic safety is not guaranteed,''
JHEP {\bf 1511}, 023 (2015).
[arXiv:1508.07411 [hep-th]].
}

\lref\MartinCR{
  S.~P.~Martin and J.~D.~Wells,
  ``Constraints on ultraviolet stable fixed points in supersymmetric gauge theories,''
Phys.\ Rev.\ D {\bf 64}, 036010 (2001).
[hep-ph/0011382].
}

\lref\BarnesJJ{
  E.~Barnes, K.~A.~Intriligator, B.~Wecht and J.~Wright,
  ``Evidence for the strongest version of the 4d a-theorem, via a-maximization along RG flows,''
Nucl.\ Phys.\ B {\bf 702}, 131 (2004).
[hep-th/0408156].
}

\lref\KutasovUX{
  D.~Kutasov,
  ``New results on the 'a theorem' in four-dimensional supersymmetric field theory,''
[hep-th/0312098].
}

\lref\BondSUY{
  A.~D.~Bond and D.~F.~Litim,
  ``Asymptotic safety guaranteed in supersymmetry,''
Phys.\ Rev.\ Lett.\  {\bf 119}, no. 21, 211601 (2017).
[arXiv:1709.06953 [hep-th]].
}

\lref\KutasovSS{
  D.~Kutasov, A.~Schwimmer and N.~Seiberg,
  ``Chiral rings, singularity theory and electric - magnetic duality,''
Nucl.\ Phys.\ B {\bf 459}, 455 (1996).
[hep-th/9510222].
}

\lref\AmaritiWC{
  A.~Amariti and K.~Intriligator,
  ``(Delta a) curiosities in some 4d susy RG flows,''
JHEP {\bf 1211}, 108 (2012).
[arXiv:1209.4311 [hep-th]].
}

\lref\KutasovIY{
  D.~Kutasov, A.~Parnachev and D.~A.~Sahakyan,
  ``Central charges and U(1)(R) symmetries in N=1 superYang-Mills,''
JHEP {\bf 0311}, 013 (2003).
[hep-th/0308071].
}
\lref\HookFP{
  A.~Hook,
  ``A Test for emergent dynamics,''
JHEP {\bf 1207}, 040 (2012).
[arXiv:1204.4466 [hep-th]].
}
\lref\DeserYX{
  S.~Deser and A.~Schwimmer,
  ``Geometric classification of conformal anomalies in arbitrary dimensions,''
Phys.\ Lett.\ B {\bf 309}, 279 (1993).
[hep-th/9302047].
}
\lref\IntriligatorJJ{
  K.~A.~Intriligator and B.~Wecht,
  ``The Exact superconformal R symmetry maximizes a,''
Nucl.\ Phys.\ B {\bf 667}, 183 (2003).
[hep-th/0304128].
}
\lref\NovikovUC{
  V.~A.~Novikov, M.~A.~Shifman, A.~I.~Vainshtein and V.~I.~Zakharov,
  ``Exact Gell-Mann-Low Function of Supersymmetric Yang-Mills Theories from Instanton Calculus,''
Nucl.\ Phys.\ B {\bf 229}, 381 (1983)..
}
\lref\ZamolodchikovGT{
  A.~B.~Zamolodchikov,
  ``Irreversibility of the Flux of the Renormalization Group in a 2D Field Theory,''
JETP Lett.\  {\bf 43}, 730 (1986), [Pisma Zh.\ Eksp.\ Teor.\ Fiz.\  {\bf 43}, 565 (1986)]..
}
\lref\ErkalSH{
  D.~Erkal and D.~Kutasov,
  ``a-Maximization, Global Symmetries and RG Flows,''
[arXiv:1007.2176 [hep-th]].
}
\lref\KutasovVE{
  D.~Kutasov,
  ``A Comment on duality in N=1 supersymmetric nonAbelian gauge theories,''
Phys.\ Lett.\ B {\bf 351}, 230 (1995).
[hep-th/9503086].
}
\lref\IntriligatorSM{
  K.~A.~Intriligator and N.~Seiberg,
  ``Phases of N=1 supersymmetric gauge theories in four-dimensions,''
Nucl.\ Phys.\ B {\bf 431}, 551 (1994).
[hep-th/9408155].
}

\lref\MachacekTZ{
  M.~E.~Machacek and M.~T.~Vaughn,
  ``Two Loop Renormalization Group Equations in a General Quantum Field Theory. 1. Wave Function Renormalization,''
Nucl.\ Phys.\ B {\bf 222}, 83 (1983)..
}

\lref\MachacekFI{
  M.~E.~Machacek and M.~T.~Vaughn,
  ``Two Loop Renormalization Group Equations in a General Quantum Field Theory. 2. Yukawa Couplings,''
Nucl.\ Phys.\ B {\bf 236}, 221 (1984)..
}

\lref\TeraoJM{
  H.~Terao and A.~Tsuchiya,
  ``Conformal dynamics in gauge theories via non-perturbative renormalization group,''
[arXiv:0704.3659 [hep-ph]].
}
\lref\LitimUCA{
  D.~F.~Litim and F.~Sannino,
  ``Asymptotic safety guaranteed,''
JHEP {\bf 1412}, 178 (2014).
[arXiv:1406.2337 [hep-th]].
}

\lref\MachacekZW{
  M.~E.~Machacek and M.~T.~Vaughn,
  ``Two Loop Renormalization Group Equations in a General Quantum Field Theory. 3. Scalar Quartic Couplings,''
Nucl.\ Phys.\ B {\bf 249}, 70 (1985)..
}

 \Title{
} {\vbox{ \centerline{Comments on asymptotic safety in four--dimensional}
\bigskip
\centerline{ $\N = 1$ supersymmetric gauge theories} }}

\bigskip
\centerline{\it Meseret Asrat }
\bigskip
\smallskip
\centerline{EFI and Department of Physics, University of
Chicago} \centerline{5640 S. Ellis Av., Chicago, IL 60637, USA }
\smallskip

\vglue .3cm

\bigskip

\bigskip
\noindent

We discuss asymptotic safety in four--dimensional $\N = 1$ supersymmetric gauge theories. Our main conclusion is that in a reasonable definition of this phenomenon there are currently no controlled examples of it; there are reasons to believe that it does not occur.

\bigskip

\Date{5/18}

 
\newsec{Introduction}

In recent years there has been a resurgence of work on asymptotic safety. Much of the interest has been in applications to quantum gravity. There has also been work on the possible occurrence of this phenomenon in four dimensional $\N = 1$ supersymmetric gauge theories. Some of this work concluded that this phenomenon occurs and is exhibited in certain models, in some cases at parametrically weak coupling.

In this paper we will revisit these models and argue in these models, and more generally, in four--dimensional $\N = 1$ supersymmetric gauge theories that the phenomenon of asymptotic safety does not occur. The reasons that asymptotic safety does not occur in some cases involve the question what exactly is meant by asymptotic safety, therefore in the following we begin with a brief general discussion of the idea of asymptotic safety in quantum field theory (QFT). 

A reasonable definition, which we will adopt here, is that an asymptotically safe QFT is one that:
\item{(1)} approaches at short distances an interacting fixed point;
\item{(2)} does not fall into the Wilsonian paradigm. 

\noindent
Item (1) is self explanatory, but (2) requires some further explanation, especially since some of the examples  below do not satisfy it. 

For the purpose of our discussion, the Wilsonian paradigm is the idea that a mathematically complete QFT is defined by specifying its ultraviolet limit, a conformal field theory (CFT), and a finite set of parameters that describe the RG flow to longer distances. These parameters can correspond to (marginally) relevant couplings in the theory, and/or expectation values of operators in the CFT. A theory that does not fall into this paradigm can be  one that is defined in terms of its infrared CFT, and a flow up the RG.

The reason that item (2) in the definition of an asymptotically safe theory is (we believe) necessary is that without it, the phenomenon in question is empty---it is part of the standard picture of Wilsonian RG, and it is not clear why it deserves a special name. 

To illustrate what we mean, consider the case of QCD, a gauge theory with gauge group $SU(N_c)$ and $N_f$ (massless) flavors of fermions (quarks) in the fundamental representation of the gauge group.  As is well known, for $N_f<11N_c/2$ the theory is asymptotically free, thus at short distances the gauge coupling goes to zero. The long distance dynamics depends on $N_f$; in particular  \BanksNN, for $N_f$ slightly below\foot{More precisely, for $N_f/N_c$ slightly below $11/2$. This is easiest to achieve when $N_f$, $N_c$ are large.}  $11N_c/2$, the theory flows at long distances into an interacting CFT, known as the Banks--Zaks (BZ) fixed point.  As $N_f$ decreases, this CFT becomes more strongly coupled, and below some critical value of $N_f$, $N_f^{(\rm cr)}$, the theory is believed to spontaneously break the chiral $SU(N_f)_L\times SU(N_f)_R$ symmetry to its diagonal subgroup, and approach in the infrared a free theory of the Goldstone bosons of the broken $SU(N_f)$. The  value of $N_f^{(\rm cr)}$ is not known in general. 

If we removed item (2) in the definition of asymptotic safety above, a simple way to construct asymptotically safe theories would be the following. Start with an asymptotically free theory with $N_f$ slightly below $11N_c/2$. This theory flows in the infrared into the BZ fixed point. We can take this fixed point to be the starting point of a Wilsonian analysis, \ie\ view it as an ultraviolet fixed point  that we can perturb to flow to long distances. An example of such a flow is obtained by adding a mass term to one of the $N_f$ flavors of quarks. This flow connects the BZ fixed point of the theory with $N_c$ colors and $N_f$ flavors, $BZ_{N_c, N_f}$, to $BZ_{N_c, N_f-1}$. It certainly satisfies criterion (1) for an asymptotically safe theory, as the ultraviolet fixed point,  $BZ_{N_c, N_f}$, is interacting, but we don't need a special name to describe it. 

In this particular case, the infrared fixed point, $BZ_{N_c, N_f-1}$, is typically also interacting (in some range of $N_f$, $N_c$), but it is easy to construct examples where the infrared theory is free. An example is turning on a mass term to multiple flavors, which describes the flow $BZ_{N_c, N_f}\to BZ_{N_c, N_f'}$. If $N_f>N_f^{(\rm cr)}>N_f'$, the infrared theory is free. 

Another standard example of an RG flow that satisfies (1) above but not (2) corresponds to generic points on the Coulomb branch of $N=4$ SYM with gauge group $SU(n)$ (say). At short distances the theory is described by a non--trivial four--dimensional CFT, while in the infrared it is a free $U(1)^{n-1}$ theory. 

The main point in the above discussion is that non--trivial fixed points are ubiquitous in four--dimensional QFT, as are RG flows from such points to other (free or interacting) CFT's, so for the phenomenon of asymptotic safety to be non--trivial, we need some version of criterion (2) above. 

The rest of the paper is organized as follows. In section 2 we examine the claim of \BondSUY\ that a certain model of supersymmetric gauge theories exhibits asymptotic safety at weak coupling. We point out that the RG flows discussed in \BondSUY\ are standard Wilsonian RG flows and thus do not satisfy criterion (2) above. These flows, however, do exhibit the  phenomenon that irrelevant (relevant) operators in the ultraviolet build up, as we flow toward low energies, negative (positive) anomalous dimension, and eventually become relevant (irrelevant) in the infrared. Such operators are known as dangerously irrelevant (harmlessly relevant), and are well known to be ubiquitous in four--dimensional supersymmetric gauge theories. More importantly for us, they are comfortably accommodated in the Wilsonian framework.

In section 3 we comment on some supersymmetric gauge theories discussed in \MartinCR\ and some more recent follow--ups. These theories are free at long distances. They are constructed in such a way that (at large values of couplings) they satisfy all the (algebraic) consistency constraints that an asymptotically safe theory would satisfy. The authors' presumption is therefore that these theories are described by interacting fixed points at short distances.

We will argue that there are actually reasons to believe that such fixed points do not exist. In such theories, there are non--perturbative techniques that allow one to study the dynamics at fixed points in a way that does not rely on a weak coupling expansion, but it often happens that these techniques break down when the coupling exceeds a certain critical value. The detailed results of \MartinCR\ as well as later work makes it plausible that the breakdown happens before one reaches the regime where the strongly coupled fixed points proposed by these authors appear.

Appendix A provides a brief review of dangerously irrelevant and harmlessly relevant operators in $\N = 1$ four--dimensional supersymmetric gauge theories. Technical details that we will use in the main text will be found in Appendix B.

\newsec{The model, $a$--function and RG flows}
 
 In this section we systematically analyze the pattern of RG flows in the model considered in \BondSUY\ and show that it does not exhibit asymptotic safety, that is, it does not satisfy the criterion (2) discussed in the introduction.
 
 In our analysis we will employ the generalized central charge $a$--function constructed in \KutasovUX.  It offers a simple and non--perturbative approach for analyzing RG flows. A brief review of this construction with an example is provided in Appendix B. In the following we begin our discussion with a brief description of the model.

The model is a class of $\N = 1$  supersymmetric  gauge theories with a product gauge group $SU(N_1)\times SU(N_2)$ coupled to (anti--)fundamental chiral suprfields ($\widetilde{\psi}_{\widetilde{i}}$)$\psi_i$$(i,{\widetilde{i}} = 1, \cdots, N_f)$, ($\widetilde{\chi}_{\widetilde{i}}$)$\chi_i$$(i,{\widetilde{i}} = 1, \cdots, N_f)$ and ($\widetilde{Q}_{\widetilde{i}}$)$Q_i$$(i,{\widetilde{i}} = 1, \cdots, N_Q)$, and (an)a ``(anti--)bi--fundamental" chiral superfield ($\Psi$)$\widetilde{\Psi}$.  It has $SU(N_f)_{L}\times SU(N_f)_{R}\times SU(N_Q)_{L}\times SU(N_Q)_{R} \times U(1)_{R}$ global symmetry. The field contents of the model, and the representations under which they transform are summarized as follows.
 
 $$
\vbox{\tabskip=0pt
\halign{\tabskip=0.2cm #\hfil&#\hfil&#\hfil&#\hfil&#\hfil&#\hfil&#\hfil&#\hfil&#\tabskip=0pt \cr
Chiral superfields & $SU(N_1)$&$SU(N_2)$&$SU(N_f)_L$&$SU(N_f)_R$&$SU(N_Q)_L$&$SU(N_Q)_R$\cr
$\psi_i(\widetilde{\psi}_{\widetilde{i}})(i,{\widetilde{i}} = 1, \cdots, N_f)$ & $\bf N_1(\bf \bar{N}_1)$&$\bf 1$&$\bf 1(\bf \bar{N}_f)$&$\bf N_f(\bf 1)$&$\bf 1(\bf 1)$&$\bf 1(\bf 1)$\cr
$\chi_i(\widetilde{\chi}_{\widetilde{i}})(i,{\widetilde{i}} = 1, \cdots, N_f)$ & $\bf 1(\bf 1)$&$\bf N_2(\bf \bar{N}_2)$&$\bf 1(\bf \bar{N}_f)$&$\bf N_f(\bf 1)$&$\bf 1(\bf 1)$&$\bf 1(\bf 1)$\cr
$Q_i(\widetilde{Q}_{\widetilde{i}})(i,{\widetilde{i}} = 1, \cdots, N_Q)$ & $\bf 1(\bf 1)$&$\bf N_2(\bf \bar{N}_2)$&$\bf 1(\bf 1)$&$\bf 1(\bf 1)$&$\bf 1(\bf \bar{N}_Q)$&$\bf N_Q(\bf 1)$\cr
$\Psi(\widetilde{\Psi})$ & $ \bf \bar{N}_1(\bf N_1)$&$\bf \bar{N}_2(\bf N_2)$&$\bf 1(\bf 1)$&$\bf 1(\bf 1)$&$\bf 1(\bf 1)$&$\bf 1(\bf 1)$\cr}}
$$

The authors included a Yukawa interaction of the form 
\eqn\su{
W = y\Tr[\psi\Psi\chi +\widetilde {\psi} \widetilde{\Psi}\widetilde{\chi}].
}

In the following we apply the construction of \KutasovUX\ and obtain the gradient flow equations of the generalized central charge $a$--function in the configuration space defined by the couplings of the model. The zeros of these equations determine RG flows fixed points. We find these equations following the general procedure outlined in Appendix B. This procedure involves constructing and extremizing an intermediate $a$--function.

The first step in the procedure is constructing an intermediate $a$--function. Adapting the expression $\rm{(B.2)}$ for the model, we find that the intermediate $a$--function $a(\lambda_i,R_j)$ $(i = 1, 2, y; j = \psi, \chi, Q, \Psi)$ takes the following form
\eqn\afunc{\eqalign{a(\lambda_i,R_j)=\ &2\left(N_1^2-1\right)+2\left(N_2^2-1\right)\cr
+ \ &2N_fN_1\left[3(R_\psi-1)^3-(R_\psi-1)\right]
+ \ 2N_fN_2\left[3(R_\chi-1)^3-(R_\chi-1)\right]\cr
+ \ &2N_2N_Q\left[3(R_Q-1)^3-(R_Q-1)\right]+2N_1N_2\left[3(R_\Psi-1)^3-(R_\Psi-1)\right]\cr
- \ &\lambda_1\left[N_1+N_f(R_\psi-1)+N_2(R_\Psi-1)\right]\cr
- \ &\lambda_2\left[N_2+N_f(R_\chi-1)+N_1(R_\Psi-1)+N_Q(R_Q-1)\right]\cr
- \ &\lambda_y\left(2-R_\psi-R_\chi-R_\Psi\right).
}}

The three Lagrange multipliers: $\lambda_1, \lambda_2$ and $\lambda_y$, are related to the two gauge couplings corresponding to the gauge groups $SU(N_1)$ and $SU(N_2)$, \ie\ $\alpha_1 := {N_1g^2_1\over(4\pi)^2}$ and $\alpha_2 := {N_2g^2_2\over(4\pi)^2}$ respectively, and the Yukawa coupling $\alpha_y := {N_1y^2\over(4\pi)^2}$. The precise relations between the multipliers and couplings, in the large $N_1, N_2$ limits, are given by \refs{\KutasovUX\BarnesJJ-\KutasovXU} 
\eqn\coupling{\eqalign{\lambda_1 = 8N_1\alpha_1 + \O(\alpha^2), \qquad \lambda_2 = 8N_2\alpha_2 + \O(\alpha^2),\qquad \lambda_y  = 8N_2N_f\alpha_y + \O(\alpha^2).
}}
Where $\alpha$ denotes any linear combination of the couplings $\alpha_i \ (i = 1, 2, y)$.

The next step is finding the local maximum of the intermediate $a$--function $a(\lambda_i,R_j)$ with respect to the unknowns $R_k$. That is, we evaluate $\partial_{R_i}a(\lambda_j, R_k) = 0$, which gives a set of equations that can be solved for $R_i := R_i(\lambda_j)$. In a certain finite region in the space of couplings around the origin, \ie\ the free fixed point $\lambda_i = 0$, we find
\eqn\finrr{\eqalign{
R_\psi= \ &1-{1\over3}\left(1+{\lambda_1N_f-\lambda_y\over 2N_fN_1}\right)^{\half},\cr
R_\chi= \ &1-{1\over3}\left(1+{\lambda_2N_f-\lambda_y\over 2N_fN_2}\right)^{\half},\cr
R_Q=\ &1-{1\over3}\left(1+{\lambda_2\over 2N_2}\right)^{\half},\cr
R_\Psi= \ &1-{1\over3}\left(1+{\lambda_1N_2+\lambda_2N_1-\lambda_y\over 2N_1N_2}\right)^{\half}.
}}

We substitute these expressions for $R_i$ in $a(\lambda_i,R_j)$ \afunc. This finally gives the generalized central charge $a$--function $a(\lambda_i):=a(\lambda_i, R_j(\lambda_k))$. Since it solves the equation $\partial_{R_l} a(\lambda_i, R_j(\lambda_k)) = 0$, its gradient flow equations in the configuration space of the model are given by
\eqn\ddaa{\eqalign{
{da(\lambda_i)\over d\lambda_1} =&-\left[N_1+N_f(R_\psi-1)+N_2(R_\Psi-1)\right],\cr
{da(\lambda_i)\over d\lambda_2} =&-\left[N_2+N_f(R_\chi-1)+N_1(R_\Psi-1)+N_Q(R_Q-1)\right],\cr
{da(\lambda_i)\over d\lambda_y} =&-\left(2-R_\psi-R_\chi-R_\Psi\right),
}}
here the running $R$--charges $R_i$ in \ddaa\ are given by their form \finrr\ in terms of the running couplings $\lambda_i$---in some renormalization scheme. 

These equations determine the fixed points of the model in a certain finite region in the space of couplings around $\lambda_i = 0$. In this paper, however, we work at weak coupling, that is, in some small region around $\lambda_i = 0$. For simplicity, we also work in the Veneziano large $N$ limit\foot{$N$ is some linear combination of $N_1, N_2, N_f, N_Q$.}, that is, we take $N_1, N_2, N_f, N_Q$ to infinity while keeping their ratios fixed. This is compatible with the weak coupling regime; this will become more evident in a later subsection.\foot{Basically, the idea is that, as in the Banks--Zaks analysis \BanksNN, the couplings can be taken to be arbitrarily small independently.} The generalization to finite $N$ is trivial. 

Therefore, we can expand the running $R$--charges $R_j$ in \ddaa\ (given by \finrr) to linear order in the couplings $\lambda_i$. Expanding \finrr\ to leading order in $\lambda_i$ yields

\eqn\expand{\eqalign{
R_\psi=\ &{2\over3}-{\lambda_1N_f-\lambda_y\over 12N_fN_1},\cr
R_\chi=\ &{2\over3}-{\lambda_2N_f-\lambda_y\over 12N_fN_2},\cr
R_Q=\ &{2\over3}-{\lambda_2\over 12N_2},\cr
R_\Psi=\ &{2\over3}-{\lambda_1N_2+\lambda_2N_1-\lambda_y\over 12N_1N_2}.
}}
Substituting these in \ddaa\ gives the linearized form of the gradient flow equations
\eqn\newdao{12N_1\frac{da}{d\lambda_1} = -\left[4N_1(3N_1-N_f-N_2) - \lambda_1(N_f+N_2)- \lambda_2N_1+2\lambda_y\right],
}
\eqn\newdatw{12N_2\frac{da}{d\lambda_2} = -\left[4N_2(3N_2-N_f-N_1-N_Q) - \lambda_1N_2 - \lambda_2(N_f+N_1+N_Q)+2\lambda_y\right],
}
\eqn\newdatt{12N_1N_2N_f\frac{da}{d\lambda_y} = -\left[2\lambda_1N_2N_f+2\lambda_2N_1N_f - \lambda_y(N_1+N_2+N_f)\right].
}

A vanishing condition on these equations, accompanied by the positivity constraint on the couplings \coupling, specifies the fixed points of the various RG flows in the coupling space of the model. In the following subsections, we consider the cases in which we turn on only one of the gauge couplings, and the case in which we turn on both the gauge couplings; these initiate RG flows in the configuration space. We will find all the resulting RG fixed points, and we will examine the admissible RG flows that interpolate among them.  

For convenience, we define the parameters $\epsilon_1$ and $\epsilon_2$ via
\eqn\num{N_2 :=  3 N_1 - N_f -  \epsilon_1,
}
\eqn\quark{N_Q :=  8N_1 - 4N_f -3\epsilon_1  -\epsilon_2.
}
Note that $\epsilon_1$ and $\epsilon_2$ are integers, and in the Veneziano large $N$ limit, they are small, that is to say, $|\epsilon_1|, |\epsilon_2| \ll N$. In this limit, the couplings $\alpha_i$ are in the weak coupling regime, \ie\ they are of $\O\left(1/N\right)$, therefore, as in the Banks--Zaks analysis \BanksNN, weak coupling (perturbative) Banks--Zaks expansion in $\epsilon_1, \epsilon_2$ is applicable; we show this as we progress. We note from \newdao\ and \newdatw\ that near the Gaussian fixed point, \ie\ $\lambda_i = 0$, for both gauge groups to be asymptotically free, \ie\ $\frac{da}{d\lambda_i} < 0$ $({\rm here} \ i = 1, 2)$, we must require both $\epsilon_1$ and $\epsilon_2$ to be positive.

For further convenience, we define a positive real number $x$ as the ratio
\eqn\ratio{x := {N_f\over N_1}.
}
The condition that the number of quarks $N_Q$, given by \quark, is positive reads now
\eqn\condonee{x < 2.
}
Note that, in this regime, the degree of the gauge group $SU(N_2)$, given by \num, is trivially positive.

In each of the following subsections, we investigate the RG flows of the model in the cases in which near the Gaussian fixed point at least one of the gauge couplings is asymptotically free:
\item{(1)} $\epsilon_1 > 0, \ \epsilon_2 > 0$; 
\item{(2)}$\epsilon_1 > 0, \ \epsilon_2 < 0$;
\item{(3)}$\epsilon_1 < 0, \ \epsilon_2 > 0$;

\noindent
and/or in the case in which neither of the gauge couplings are asymptotically free
\item{(4)} $\epsilon_1 < 0, \ \epsilon_2 < 0$.


\subsec{Turning the gauge coupling $\lambda_1$ on}
In this subsection we assume that at the Gaussian fixed point the gauge coupling $\lambda_1 \approx \alpha_1$ is marginally relevant, that is, $\epsilon_1 > 0$. The gauge coupling $\lambda_2 \approx \alpha_2$ can be marginally relevant or irrelevant. We will investigate these possibilities here in the subsequent several cases. However, for the moment, we only focus on turning the gauge coupling $\lambda_1$ on.

Turning $\lambda_1$ on results in an RG flow toward long distances. The resulting fixed point of this flow is obtained by setting \newdao\ to zero with $\lambda_2 = 0$ and $\lambda_y = 0$. Note that since we are turning only $\lambda_1$ on, we do not impose the vanishing condition on \newdatw\ and \newdatt. One finds in the infrared the Banks--Zaks fixed point $(\lambda_1^*, 0, 0)$. Here,
\eqn\newdaoz{\lambda_1^* = \frac{4}{3}\epsilon_1 .
}
 
Note that this fixed point is the fixed point $BZ_1$ in \BondSUY. To see this, we first note that the two sets of parameters, \ie\ $\epsilon$, $P$ defined in \BondSUY\ and $\epsilon_1$, $\epsilon_2$ defined here, are related as
 \eqn\para{N_1\epsilon = - \epsilon_1, \qquad N_2P\epsilon = - \epsilon_2.
}
The desired result then follows readily using this and \coupling. 

We note also using \newdaoz\ that the gauge coupling $\alpha_1 \approx \lambda_1/N_1$ is of $\O\left(1/N\right)$, therefore, in the large $N$ limit, it is in the weak coupling regime. This is consistent with our earlier assumption. One can also check that this is indeed the situation in all the other cases that we will consider below. 

We note from \newdatt\ with $\lambda_2 = 0$ and $\lambda_y = 0$ that the  Yukawa coupling $\lambda_y \approx \alpha_y$ is relevant at the fixed point \newdaoz\ on the $\lambda_1$ axis; it is a dangerously irrelevant coupling. Therefore, one can turn it on at this fixed point. Turning on the Yukawa coupling drives the theory further down in the infrared into the fixed point $(\lambda_1^*, 0, \lambda_y^*)$. We find from \newdao\ and \newdatt\ with $\lambda_2 = 0$ that
 \eqn\finonde{\lambda_1^{*} = {4\over{\left(x - \frac{3}{2}\right)^2 + \frac{3}{4}}}\epsilon_1, \qquad \lambda_y^{*} = N_1\cdot {2x(3 - x)\over{\left(x - \frac{3}{2}\right)^2 + \frac{3}{4}}}\epsilon_1.
}
Note that this fixed point is the fixed point $GY_1$ in \BondSUY. One can see this easily using \coupling\ and \para\ in combination with \num\ which, in terms of the parameters defined in \BondSUY, is
\eqn\xx{
 x = \epsilon - R + 3. 
}

In the following we investigate whether $\lambda_2$ is relevant at the fixed points \newdaoz\ along the $\lambda_1$ axis and \finonde\ on the $\lambda_2 = 0$ plane.  In the case(s) in which it is relevant we can turn it on, that is to say, we can use it to probe other fixed point(s) in the infrared. There are two cases:
\bigskip
\item{(1)} $\epsilon_2 > 0$
\bigskip
We first examine the case in which $\epsilon_2$ is positive, that is, at the Gaussian fixed point the gauge coupling $\lambda_2$ is marginally relevant. 

We first consider the RG flow that resulted the fixed point \newdaoz\ on the $\lambda_1$ axis.  It follows from \newdatw\ with $\lambda_2 = 0$ and $\lambda_y = 0$ that the gauge coupling $\lambda_2$ remains relevant\foot{To be precise, it is relevant only away from the free fixed point. At the fixed point, it is marginal.} along this flow provided the constraint
 \eqn\finonec{  \lambda_1 < 4\epsilon_2,
}
is satisfied. Along this flow, $\lambda_1 \leq \lambda_1^*$; thus, the gauge coupling $\lambda_2$ is relevant throughout the flow (away from the free fixed point) and in particular at the fixed point \newdaoz\ along the $\lambda_1$ axis only if the constraint
\eqn\finoneex{{\epsilon_1\over \epsilon_2} < 3,
}
is satisfied. 

Thus, if the constraint \finoneex\ is not satisfied, then what happens is that as we flow toward low energies, the gauge coupling $\lambda_2$ that was initially marginally relevant in the ultraviolet develops positive anomalous dimension, and eventually becomes irrelevant in the infrared. Therefore, in the case in which \finoneex\ is not satisfied the gauge coupling $\lambda_2$ is an example of a harmlessly relevant coupling (see Appendix A for a brief review and examples of this class of operators).

In the regime \finoneex\ we can turn on $\lambda_2$ at the fixed point \newdaoz\ along the $\lambda_1$ axis, and depending on whether or not a condition is satisfied, the theory flows either into the fixed point $(2.24)$ (see below) on the $\lambda_2$ axis or into the fixed point $(2.32)$ (see below) on the $\lambda_y = 0$ plane. We will come back to this later.

We next consider the RG flow that was generated by turning the Yukawa coupling $\lambda_y$ on at the fixed point \newdaoz\ on the $\lambda_1$ axis. We ask the question that to be able to turn on $\lambda_2$ at the fixed point \finonde\ on the $\lambda_2 = 0$ plane, that is, to flow further to longer distances, what are the constraints that $\epsilon_1$ and $\epsilon_2$, and/or $x$ must satisfy?

 We first consider the case in which the constraint \finoneex\ is not satisfied, that is, $\lambda_2$ is harmlessly relevant and therefore irrelevant at the fixed point \newdaoz\ on the $\lambda_1$ axis. There are two cases: $x< 1$ and $x\geq 1$. 
 
 First we consider the case $x < 1$. It follows from \newdatw\ (and the assumption that \finoneex\ is not satisfied) that $x$ must satisfy the constraint 
\eqn\ifthenonex{x_0 < x < 1,
}
here $x_0$ is defined via the equation 
\eqn\ifthentwocdsexx{{\epsilon_1\over \epsilon_2} \ = \ r_2(x_0) := -{\left(x_0 - {3\over2}\right)^2 +{3\over 4}\over x_0 - 1}.
}

Therefore, in the regime \ifthenonex, $\lambda_2$ becomes once again relevant as we flow toward the fixed point \finonde\ on the $\lambda_2 = 0$ plane, thus (in this flow) it is dangerously irrelevant; one can turn it on at the fixed point \finonde\ on the plane $\lambda_2 = 0$, and flow further down in the infrared into the fixed point, see below,  ($\lambda_1^*, \lambda_2^*, \lambda_y^*$). 

For $x \geq 1$, on the other hand, we find regardless of $\epsilon_1, \epsilon_2$ that the coupling $\lambda_2$ is relevant at the fixed point \finonde\ on the $\lambda_2 = 0$ plane, and thus, in this case also one can turn it on.

In the regime \finoneex, we find that the gauge coupling $\lambda_2$ is relevant at the fixed point \finonde\ on the $\lambda_2 = 0$ plane with no constraints on the parameters of the model. 

\bigskip
\item{(2)} $ \epsilon_2 < 0$
\bigskip
In this case we take at the Gaussian fixed point the gauge coupling $\lambda_2$ to be marginally irrelevant. 

We first consider the RG flow that resulted the fixed point \newdaoz\ on the $\lambda_1$ axis. It follows from \newdatw\ (or equivalently from \finonec) with $\lambda_2 = 0$ and $\lambda_y = 0$ that irrespective of the values of $|\epsilon_2|$, the gauge coupling $\lambda_2$ remains irrelevant along this flow (away from the free fixed point), and in particular it is irrelevant at the resulting fixed point \newdaoz\ on the $\lambda_1$ axis. 

We next consider the RG flow that was generated by turning the Yukawa coupling $\lambda_y$ on at the fixed point \newdaoz\ on the $\lambda_1$ axis. There are two cases: $x > 1$ and $x \leq 1$. 

First we consider $x > 1$. We note using \newdatw\ that as we flow toward the fixed point \finonde\ on the plane $\lambda_2 = 0$, the gauge coupling $\lambda_2$ can become relevant provided the constraints 
\eqn\ifthentwox{ \left|{\epsilon_1\over \epsilon_2}\right| > 1,\qquad x_0 < x < 2,
}
are satisfied. Here $x_0$ is defined via the equation 
\eqn\ifthentwocxx{\left|{\epsilon_1\over \epsilon_2}\right| = -r_2(x_0).
}

Therefore, in this regime, the coupling $\lambda_2$ is a dangerously irrelevant coupling. Thus, it is relevant at the fixed point \finonde\ on the $\lambda_2 = 0$ plane, and it can be turned on to flow further into the fixed point, see below, ($\lambda_1^*, \lambda_2^*, \lambda_y^*$) in the infrared.

In this case, \ie\ $\epsilon_1 > 0, \ \epsilon_2 < 0$, and in the regime \ifthentwox, the authors of \BondSUY\ call the fixed point \finonde\ on the $\lambda_2 = 0$ plane at which the dangerously irrelevant coupling $\lambda_2$ is relevant (and similarly as we will see shortly in the following subsection, in the case $\epsilon_1 < 0, \ \epsilon_2 > 0$, the authors call the fixed point on the $\lambda_1 = 0$ plane at which the dangerously irrelevant coupling $\lambda_1$ is relevant) an asymptotically safe interacting ultraviolet fixed point.  As we explained in the introduction, none of these flows, however, imply asymptotic safety but rather they merely correspond to the RG flow phenomenon in which an initially irrelevant coupling at relatively high energies develops, as we flow, negative anomalous dimension and eventually becomes relevant at low energies.

For $x \leq 1$, on the other hand, regardless of the values of $|\epsilon_2|$, the gauge coupling $\lambda_2$ is irrelevant at the fixed point \finonde\ on the plane $\lambda_2 = 0$. Therefore, it switches off in the infrared.

\subsec{Turning the gauge coupling $\lambda_2$ on}

In this subsection we assume that at the Gaussian fixed point the gauge coupling $\lambda_2$ is marginally relevant, that is, $\epsilon_2 > 0$. The gauge coupling $\lambda_1$ can be marginally relevant or irrelevant. We will consider these possibilities shortly. 

Turning $\lambda_2$ on drives the theory to flow in the infrared into the Banks--Zaks fixed point $(0, \lambda_2^*, 0)$, here $\lambda_2^*$ is obtained by setting \newdatw\ to zero. With $\lambda_1 = 0$ and $\lambda_y = 0$ one finds
 \eqn\finoneeex{ \lambda_2^* = \frac{4}{3}\epsilon_2.
}
This fixed point is the fixed point $BZ_2$ in \BondSUY .

We note from \newdatt\ with $\lambda_1 = 0$ and $\lambda_y = 0$ that the Yukawa coupling $\lambda_y$ is relevant at the fixed point \finoneeex\ on the $\lambda_2$ axis; it is a dangerously irrelevant coupling. If we turn on the Yukawa coupling $\lambda_y$ at this fixed point, then the theory flows into the fixed point $(0, \lambda_2^*, \lambda_y^*)$ in the infrared. With $\lambda_1 = 0$, we find by setting \newdatw\ and \newdatt\ to zero that
 \eqn\finondeq{\lambda_2^{*} = {4(3 - x)\over{9 - 4x}}\epsilon_2, \qquad \lambda_y^{*} = N_1 \cdot {2x(3 - x)\over{9 - 4x}} \epsilon_2.
}
 This is the fixed point $GY_2$ in \BondSUY. 

In the following we investigate whether $\lambda_1$ is relevant at the fixed points \finoneeex\ along the $\lambda_2$ axis and \finondeq\ on the plane $\lambda_1 = 0$.  In the case(s) in which it is relevant, we can use it to access other low energy fixed point(s). Here also we have two cases:

\bigskip
\item{(1)} $\epsilon_1 > 0$
\bigskip

We first examine the case in which $\epsilon_1$ is positive, that is, at the Gaussian fixed point the gauge coupling $\lambda_1$ is marginally relevant. 

We first consider the RG flow that resulted the fixed point \finoneeex\ on the $\lambda_2$ axis. We find using \newdao\ that for $\lambda_1$ to remain relevant along this flow (away from the free fixed point) one must demand that
 \eqn\finonecx{ \lambda_2 < 4\epsilon_1.
}
Since along the flow $\lambda_2 \leq \lambda_2^*$, $\lambda_1$ is relevant throughout the flow (away from the free fixed point) and in particular at the fixed point \finoneeex\ along the $\lambda_2$ axis provided
 \eqn\finoneees{  {\epsilon_1\over \epsilon_2} > {1\over 3}. 
}

Thus, if the constraint \finoneees\ is not satisfied, then the gauge coupling $\lambda_1$ can become irrelevant in the infrared. In the case in which the constraint \finoneees\ is not met, it is a harmlessly relevant coupling.

In the regime \finoneees\ we can turn on $\lambda_1$ at the fixed point \finoneeex\ along the $\lambda_2$ axis, and flow either into the fixed point \newdaoz\ on the $\lambda_1$ axis or into the fixed point $(2.32)$ (see below) on the $\lambda_y = 0$ plane depending on whether or not a condition is satisfied. We will comment on this later.

We next consider the RG flow that was generated by turning the Yukawa coupling $\lambda_y$ on at the fixed point \finoneeex\ on the $\lambda_2$ axis. 

Assuming the constraint \finoneees\ is not satisfied, that is, assuming that $\lambda_1$ is harmlessly relevant and thus irrelevant at the fixed point \finoneeex\ on the $\lambda_2$ axis, we now ask the question, what are the constraints that the parameters $\epsilon_1$ and $\epsilon_2$, and/or $x$ must respect so that $\lambda_1$ becomes relevant as we flow toward the fixed point $(0, \lambda_2^*, \lambda_y^*)$? There are two cases: $x < 1$ and $x\geq 1$. 

We begin with the former. We find using \newdao\ that $\lambda_1$ can become once again relevant provided the constraint
\eqn\ifthenonexxx{x_0 < x < 1,
}
is satisfied. Here $x_0$ is defined via the equation 
\eqn\ifthentwocddsexx{{\epsilon_1\over \epsilon_2}\ = \ r_1(x_0) := -{(3-x_0)(x_0-1)\over 9 - 4x_0}.
}

Therefore, in the regime \ifthenonexxx, $\lambda_1$ is dangerously irrelevant, thus, it is relevant at the fixed point \finondeq\ on the plane $\lambda_1 = 0$, and one can turn it on at this fixed point to flow further down in the infrared into the fixed point, see below,  ($\lambda_1^*, \lambda_2^*, \lambda_y^*$). 

On the other hand, we find for $x \geq 1$ that irrespective of $\epsilon_1$, $\epsilon_2$ the gauge coupling $\lambda_1$ is relevant at the fixed point \finondeq\ on the $\lambda_1 = 0$ plane.

In the regime \finoneees, the coupling $\lambda_1$ is relevant at the fixed point \finondeq\ on the $\lambda_1 = 0$ plane with no restrictions.

\bigskip
\item{(2)} $\epsilon_1 < 0$
\bigskip

In this case we take at the Gaussian fixed point the gauge coupling $\lambda_1$ to be marginally irrelevant. 

We first consider the RG flow that ended at the fixed point \finoneeex\ on the $\lambda_2$ axis. It follows from \newdao (or equivalently from \finonecx) that irrespective of the values of $|\epsilon_2|$, the gauge coupling $\lambda_2$ is irrelevant along this flow (away from the free fixed point), and in particular, it is irrelevant at the fixed point \finoneeex\ on the $\lambda_2$ axis. Therefore, it goes to zero in the infrared.

For the gauge coupling $\lambda_1$ to become relevant in the infrared one must turn on the Yukawa coupling at the fixed point \finoneeex\ on the $\lambda_2$ axis. There are two cases: $x > 1$ and $x \leq 1$. 

In the former case, the constraints that the parameters $\epsilon_1$ and $\epsilon_2$, and $x$ must satisfy so that the gauge coupling $\lambda_1$ becomes relevant in the infrared as we flow toward the fixed point \finondeq\ on the plane $\lambda_1 = 0$  are
\eqn\ifthenthreee{ \left|{\epsilon_1\over \epsilon_2}\right| < 1, \qquad x_0 < x < 2.
}
Here $x_0$ is defined via the equation 
\eqn\ifthentwocxx{\left|{\epsilon_1\over \epsilon_2}\right| = -r_1(x_0).
}

If these constraints are satisfied, then $\lambda_1$ becomes relevant as we flow toward the infrared. Therefore, in this regime, the coupling $\lambda_1$ is a dangerously irrelevant coupling. Thus, turning on $\lambda_1$ at the fixed point \finondeq\ on $\lambda_1 = 0$ plane generates a flow into the infrared fixed point, see below,  ($\lambda_1^*, \lambda_2^*, \lambda_y^*$). 

We also note here that in the regime \ifthenthreee\ the fixed point \finondeq\ does not correspond to an asymptotically safe fixed point for the same reason as in the earlier case.

For $x \leq 1$, irrespective of the values of $|\epsilon_1|$, the gauge coupling $\lambda_1$ is irrelevant at the fixed point \finondeq\ on the plane $\lambda_1 = 0$; it flows to zero in the infrared.


\subsec{Turning both the gauge couplings $\lambda_1$ and $\lambda_2$ on}

There are two cases to consider: $(1)$ $\lambda_y = 0$, and $(2)$ $\lambda_y \neq 0$. In what follows we study these cases.

We first consider turning on only the gauge couplings, \ie\ we take $\lambda_y = 0$. In this case the theory flows in the infrared into the fixed point $(\lambda_1^*, \lambda_2^*, 0)$. This fixed point is obtained by setting \newdao\ and \newdatw\ to zero with $\lambda_y = 0$. One gets
\eqn\fixedoneee{\lambda_1^{*} = \frac{1}{2}(3\epsilon_1 - \epsilon_2), \qquad \lambda_2^{*} = \frac{1}{2}(3\epsilon_2 - \epsilon_1).
}
This fixed point is the fixed point $BZ_{12}$ in \BondSUY.

On the other hand, if we also turn on the Yukawa coupling $\lambda_y$,  then the theory flows in the infrared into the fixed point $(\lambda_1^*, \lambda_2^*, \lambda_y^*)$. Setting the equations \newdao, \newdatw\  and \newdatt\ to zero we obtain
\eqn\fixedonee{\lambda_1^{*} = \frac{\epsilon_1(9 - 4x) + \epsilon_2(3-x)(x - 1)}{\frac{1}{4}(2-x)\left[3(x - \frac{5}{3})^2 + \frac{11}{3}\right]},
}
\eqn\fixedtwoo{\lambda_2^{*} = \frac{\epsilon_1(3-x)(x - 1) + \epsilon_2(3-x)\left[(x - \frac{3}{2})^2 + \frac{3}{4}\right]}{\frac{1}{4}(2-x)\left[3(x - \frac{5}{3})^2 + \frac{11}{3}\right]},
}
\eqn\fixedukawaa{\lambda_y^{*} = N_f\cdot\frac{\epsilon_1(3-x)(8 - 3x) + \epsilon_2(3-x)x}{\frac{1}{2}(2-x)\left[3(x - \frac{5}{3})^2 + \frac{11}{3}\right]} =  N_f\cdot\frac{1}{2}\left[(3 - x)\lambda_1^* + \lambda_2^*\right].
}
This fixed point is the fixed point $GY_{12}$ in \BondSUY. 

We next check that these fixed points are in the physical regime. That is, we find the regimes in the parameter space of the model in which $\lambda_1^{*}, \ \lambda_2^{*}$ (and $\lambda_y^{*}$) are positive. There are four cases: 

\bigskip
\item{(1)} $\epsilon_1 > 0, \ \epsilon_2 > 0$
\bigskip

In this case both gauge couplings are marginally relevant at the Gaussian fixed point. The fixed point \fixedoneee\ with $\lambda_y^* = 0$ is physical if the constraint \eqn\finone{ {1\over 3} \leq {\epsilon_1\over \epsilon_2} \leq 3,
}
is satisfied. We note that this constraint consists of the intersection of the constraint regions \finoneex\ and \finoneees. This make sense in that one can reach the fixed point \fixedoneee\ on the plane $\lambda_y = 0$ either by first turning on $\lambda_1$, and then $\lambda_2$ or vice versa; however, note that this is true only if the (gauge) couplings are not harmlessly relevant. We will come back to this point shortly. On either side of the constraint region \finone\ we note from \finoneex\ and \finoneees\ that either of the gauge couplings is harmlessly relevant, and thus turning on both (at the free fixed point) drives the theory to flow either into the fixed point along the $\lambda_1$ or $\lambda_2$ axis. We will comment on this shortly. We also note that the Yukawa coupling $\lambda_y$ is relevant at \fixedoneee\ on the plane $\lambda_y = 0$, and thus we can turn it on, which in turn generates a flow toward the fixed point given by \fixedonee, \fixedtwoo\ and \fixedukawaa\ in the infrared.

In subsection $(2.1)$, in the case in which $\epsilon_2 > 0$,  we saw that in the regime \finoneex\ the gauge coupling $\lambda_2$ is relevant at the fixed point \newdaoz\ on the $\lambda_1$ axis. We mentioned that in this regime, therefore, one can turn on this coupling to flow in the infrared, depending on whether or not a condition is satisfied, either into the fixed point \finoneeex\ on the $\lambda_2$ axis or into the above fixed point \fixedoneee\ on the $\lambda_y = 0$ plane.  In the following we will be very specific.

Suppose that $\epsilon_1$ and $\epsilon_2$ satisfy the constraints \finoneex\ and \finoneees. Thus, they also satisfy the constraint \finone\ since
\eqn\cononeppx{
{1\over 3} < {\epsilon_1\over \epsilon_2} < 3.
}
In this case, it follows from our earlier discussion and \finone\ that turning $\lambda_1$, and then $\lambda_2$ results in an RG flow that ends at the fixed point \fixedoneee\ on the plane $\lambda_y = 0$; note that in this regime none of the gauge couplings are harmlessly relevant throughout the flow.

Now suppose, to the contrary, that $\epsilon_1$ and $\epsilon_2$ do not satisfy \finoneees\ (and therefore \cononeppx), that is, suppose that
\eqn\cononep{
{\epsilon_1\over \epsilon_2} \leq {1\over 3}.
}
Note that only the equal sign satisfies the constraint \finone. Note also that in this regime the condition \finoneex\ is trivially satisfied, and therefore, the coupling $\lambda_2$ is relevant at the fixed point on the $\lambda_1$ axis, and as such we can turn it on at this fixed point and flow to lower energies. In this case, turning $\lambda_1$, and then $\lambda_2$ will take us into the fixed point \finoneeex\ on the $\lambda_2$ axis; therefore, also in this regime, turning both the gauge couplings $\lambda_1$ and $\lambda_2$ on (at the free fixed point) leads in the infrared into the same fixed point on the $\lambda_2$ axis. This can be understood as follows. 

Consider deforming the fixed point \newdaoz\ on the $\lambda_1$ axis by turning on the relevant gauge coupling $\lambda_2$. For a non--zero value of $\lambda_2$, we note from \newdao\ that the coupling $\lambda_1$ at this fixed point is irrelevant.  Therefore, as we flow toward lower energies, the gauge coupling $\lambda_1$ shrinks toward zero, and the gauge coupling $\lambda_2$, on the other hand, grows away from zero. In the regime \cononep, we find from \newdao\ and \newdatw, the rate at which the coupling $\lambda_1$ runs to zero is much faster than the rate at which the coupling $\lambda_2$ runs away from zero. Therefore, when $\lambda_1$ (first) hit zero, we find the fixed point \finoneeex\ at $\lambda_2 = \lambda_2^*$ on the $\lambda_2$ axis. We note from \finoneees\ that in the regime \cononep\ the coupling $\lambda_1$ is irrelevant at this fixed point, therefore it does not grow, \ie\ it stays zero. The change in the central charge $a$ along the flow can be obtained using the generalized $a$--function. We find in accord with the weak $a$--theorem $\Delta a = a(\lambda_1^*, 0,0) - a(0, \lambda_2^*,0) > 0$.

In the regime \cononep, thus, what happens when we turn both the gauge couplings $\lambda_1$ and $\lambda_2$ on at the free fixed point is that as we flow toward infrared, $\lambda_1$ develops positive anomalous dimension, and eventually becomes irrelevant---thus it is harmlessly relevant---while $\lambda_2$ stays relevant (except at the free fixed point where it is marginally relevant). Thus, the RG flow then finally ends at the fixed point \finoneeex\ on the $\lambda_2$ axis.

We also stated above in subsection $(2.2)$ in the case in which $\epsilon_1 > 0$ that in the regime \finoneees\ one can turn on the gauge coupling $\lambda_1$ at the fixed point \finoneeex\ on the $\lambda_2$ axis to flow in the infrared, depending on whether or not a condition is satisfied, either into the fixed point \newdaoz\ on the $\lambda_1$ axis or into the above fixed point \fixedoneee\ on the $\lambda_y = 0$ plane. In the following we briefly discuss these two RG flows.

If $\epsilon_1$ and $\epsilon_2$ satisfy the constraint \cononeppx, then it follows from our earlier discussion and \finone\ that turning $\lambda_1$ on at the fixed point \finoneeex\ on the $\lambda_2$ axis drives the theory into the above fixed point on the plane $\lambda_y = 0$. 

On the contrary, if  $\epsilon_1$ and $\epsilon_2$ do not satisfy \cononeppx, that is, if 
\eqn\cononep{
{\epsilon_1\over \epsilon_2} \ge 3,
}
then the theory flows into the fixed point \newdaoz\ on the $\lambda_1$ axis. Here also the change in the central charge $\Delta a =  a(0, \lambda_2^*, 0) - a(\lambda_1^*, 0, 0) > 0$.

We note from \finoneex\ that in the regime \cononep\ the coupling $\lambda_2$ is harmlessly relevant, and for essentially the same reason as in the earlier case, therefore, in this regime turning both the gauge couplings $\lambda_1$ and $\lambda_2$ on at the free fixed point leads in the infrared into the fixed point \newdaoz\ on the $\lambda_1$ axis. 

We next examine the fixed point given by \fixedonee, \fixedtwoo\ and \fixedukawaa. There are four cases, and we discuss them in turn in the following.

The first case is when the parameters $\epsilon_1$ and $\epsilon_2$ are allowed to take any values. In this case we find that this fixed point is physical if the constraint 
\eqn\caseoneone{1\leq x<2,
}
is satisfied. This is consistent with the results in subsections $2.1$ and $2.2$.

The next case is when the parameters $\epsilon_1$ and $\epsilon_2$ are such that
\eqn\finonex{ {1\over 3} \leq {\epsilon_1\over \epsilon_2} \leq 3.
}
In this case we find that the constraint is 
\eqn\caseonetwodf{0 < x <2.
}
We note from \condonee\ that this constraint is trivially satisfied. That is, as long as we are in the regime \finonex, there is always a physical infrared fixed point that is given by \fixedonee, \fixedtwoo\ and \fixedukawaa.

The other case is when 
\eqn\finoneiox{ {\epsilon_1\over \epsilon_2} > 3.
}
In this case the constraint on $x$ is
\eqn\caseonetwo{x_0 < x<1,
}
here $x_0$ is defined via the equation
\eqn\ifthenone{  {\epsilon_1\over \epsilon_2} = r_2(x_0).
}

The final case is when 
\eqn\finoneiox{ {\epsilon_1\over \epsilon_2} < {1\over3}.
}
In this case the constraint is
\eqn\caseonioetwo{x_0 < x<1,
}
here $x_0$ is defined via the equation
\eqn\iftheionone{  {\epsilon_1\over \epsilon_2} = r_1(x_0).
}
\iffalse
On the other hand, for any $\epsilon_1, \epsilon_2$, the fixed point given by \fixedonee, \fixedtwoo\ and \fixedukawaa\ is physical if the constraint
\eqn\caseoneone{1\leq x<2,
}
is satisfied. This is consistent with \ifthenonex\ and \ifthenonexxx.

In general, however, the constraint is 
\eqn\caseonetwo{x_0 < x<2,
}

here $x_0 = {\rm {max}} \{x_1, x_2 \}$, and $x_1$ and $x_2$ are some positive (real) numbers that are defined via  
\eqn\ifthenone{  {\epsilon_1\over \epsilon_2} = r_1(x_1), \qquad {\epsilon_1\over \epsilon_2} = r_2(x_2).
}
When the parameters $\epsilon_1$ and $\epsilon_2$ are such that
\eqn\finonex{ {1\over 3} < {\epsilon_1\over \epsilon_2} < 3,
}
$x_1$ and $x_2$ are defined to take the same value $0$. Thus, in the regime \finonex, the constraint is
 \eqn\caseonetwodf{0 < x <2.
}
We note from \condonee\ that this constraint is trivially satisfied. That is, as long as we are in the regime \finonex, there is always a physical infrared fixed point that is given by \fixedonee, \fixedtwoo, and \fixedukawaa.
\fi

Note that although the constraint \finone\ is the same as \finonex, the former is associated with a different fixed point that lies on the $\lambda_y = 0$ plane. We also note that the constraint \finone\ being identical to \finonex\ is consistent with the fact that one can turn on the Yukawa coupling $\lambda_y$ at the fixed point \fixedoneee\ on the $\lambda_y = 0$ plane, and flow into the fixed point given by \fixedonee, \fixedtwoo\ and \fixedukawaa. Note also that the constraint \caseonetwo\ is the constraint \ifthenonex, and the constraint \caseonioetwo\ is the constraint \ifthenonexxx.

\bigskip
\item{(2)} $\epsilon_1 > 0,\ \epsilon_2 < 0 $
\bigskip

In this case at the free fixed point $\lambda_1$ is marginally relevant, and $\lambda_2$ is marginally irrelevant. Since $\lambda_2^*$ in \fixedoneee\ is negative, the fixed point given by \fixedoneee\ on the $\lambda_y = 0$ plane is unphysical.

Now consider the fixed point given by \fixedonee, \fixedtwoo\ and \fixedukawaa. Note that to have a physical fixed point one must allow $x$ to take values only in the range $1 < x < 2$ (since for $x \leq 1$, $\lambda_2^* < 0$). We find that for 
\eqn\casetwoone{ \left|{\epsilon_1\over \epsilon_2}\right| > 1,\qquad x_0 < x<2,
}
this fixed point is physical. Here $x_0$ is defined via the equation
\eqn\ifthentwo{\left|{\epsilon_1\over \epsilon_2}\right| =  -r_2(x_0). 
}
 We note that these constraints are exactly the constraints \ifthentwox. This is the case because, since $\lambda_2$ is not dangerously irrelevant, it switches off as we flow toward the infrared, and thus we are effectively turning on only $\lambda_1$.

\bigskip
\item{(3)} $\epsilon_1 < 0,\ \epsilon_2 > 0$
\bigskip

Here at the free fixed point the gauge coupling $\lambda_1$ is marginally irrelevant and the gauge coupling $\lambda_2$ is marginally relevant. Since $\lambda_1^*$ in \fixedoneee\ is negative, the fixed point given by \fixedoneee\ on the $\lambda_y = 0$ plane is also, as in the previous case, unphysical.

Consider the fixed point given by \fixedonee, \fixedtwoo\ and \fixedukawaa. We note that to have a physical fixed point one must restrict $x$ in the range $1 < x < 2$ (since for $x \leq1$, $\lambda_1^* <0$).  We find that for 
\eqn\casethreeone{ \left|{\epsilon_1\over \epsilon_2}\right| < 1,\qquad x_0 < x<2,
}
this fixed point is physical. Here $x_0$ is defined via the equation
\eqn\ifthenthree{\left|{\epsilon_1\over \epsilon_2}\right| =  -r_1(x_0). 
}
We note also here that these constraints are precisely the constraints \ifthenthreee. The reason is due to the fact that $\lambda_1$ is not dangerously irrelevant, and thus it flows to zero in the infrared.

\bigskip
\item{(4)} $\epsilon_1 < 0,\ \epsilon_2 < 0 $
\bigskip

So far in the two preceding subsections, we discussed the cases in which at least one of the gauge groups is asymptotically free. Here, however, we consider the case where both $\epsilon_1$ and $\epsilon_2$ are negative. That is, at the Gaussian fixed point both gauge couplings are marginally irrelevant, thus, none of the gauge groups are asymptotically free. 

From \fixedoneee\ we find that for both $\lambda_1^*$ and $\lambda_2^*$ to be positive we must require both the conditions $\left|\epsilon_1/\epsilon_2\right| < 1/3$ and $\left|\epsilon_1/\epsilon_2\right| > 3$. However, these conditions cannot be met simultaneously. Therefore, the fixed point \fixedoneee\ on the $\lambda_y = 0$ plane is unphysical. 

We also note that $\lambda_y^*$ in \fixedukawaa\ is unphysical when both $\epsilon_1$ and $\epsilon_2$ are negative, hence the fixed point given by \fixedonee, \fixedtwoo\ and \fixedukawaa\ is unphysical as well. Therefore, there are no interacting ultraviolet fixed points of the Yukawa and gauge couplings. 

We argue in the next section more generally that it is not possible to achieve asymptotic safety in four dimensional $\N = 1$ supersymmetric gauge theories with or without superpotentials, and in particular, at weak coupling.


\newsec{Discussion}

In this section we comment on certain supersymmetric gauge theories considered in \MartinCR\ and some more recent follow--ups. 

The authors argued that asymptotic safety is realizable and it exhibits certain theories. These theories satisfy all the consistency constraints that one would require an asymptotically safe theory to satisfy. These constraints are algebraic in nature. They include the constraints that at fixed points the scaling dimensions of gauge invariant operators cannot be less than one and the central charges defined via two point functions of the stress tensor and conserved currents must be positive, and that the Euler anomaly coefficient $a$ must be positive at fixed points and satisfy the inequality $\Delta a := a_{{\rm UV}} - a_{{\rm IR}}>0$, which is known as the weak $a$--theorem. The weak $a$--theorem constraint is equivalent to the constraint that at least one of the chiral superfields in the theory must have an $R$--charge larger than $5/3$.

An example given is an $SU(N_c)$ gauge theory with $N_f$ pairs of fundamentals and anti--fundamentals $Q, \tilde{Q}$, two adjoints $A_1, A_2$ and a superpotential 
\eqn\potas{W  = y_1A_1Q\tilde{Q} + y_2A_1^3,
} that does not involve the adjoint $A_2$. 

In the regime $N_f < N_c$ this theory is free in the ultraviolet, and in the infrared it flows into a non--trivial fixed point. The $R$--charges of the superfields at this fixed point are 
\eqn\asr{R_{Q} = R_{\tilde{Q}} = \frac{2}{3},\quad R_{A_1} = \frac{2}{3},\quad  R_{A_2} = \frac{N_c + N_f}{3N_c}.
}

In the regime $N_f > N_c$ this theory is infrared free. The existence of an interacting ultraviolet fixed point in a certain finite region around the infrared free fixed point can be analyzed using the recent generalized central charge $a$--function construction \KutasovUX\ (see Appendix B). At weak coupling with $N_f$ very close to $N_c$, \ie\ $N_f/N_c = 1 + \epsilon, \ 0 < \epsilon \ll 1$, we find that there is no fixed point in the ultraviolet. The authors took $N_f$ very far away from $N_c$, \ie\ $N_f > 4N_c$, so that at least one of the $R$--charges is large enough to satisfy the above $5/3$ constraint. For $N_f > 4N_c$, therefore all the constraints are satisfied.

We next argue more generally that the  constraint that at least one of the $R$--charges must be larger than $5/3$ cannot be satisfied in four dimensional $\N = 1$ supersymmetric gauge theories in a regime in which the original infrared degrees of freedom, that is to say, variables, do not break down.

Without loss of generality, we consider a supersymmetric gauge theory with a gauge group $G$ and coupling $g$ containing chiral superfields $\Phi_i$ in irreducible representations $r_i$. We take a superpotential of the form 
\eqn\pot{W =y\prod_i\left(\Phi_i\right)^{n_i},
}
where $n_i$ are some positive integers. The exact running $R$--charges $R_i$ of $\Phi_i$ along an RG flow are given by $\rm{(B.3)}$ in Appendix B. For convenience we write it here again
\eqn\rnmx{R_i(\lambda, \lambda_{y}) = 1 - \frac{1}{3}\left(1 + \frac{\lambda T(r_i) - \lambda_{y}n_{i}}{|r_i|}\right)^{\frac{1}{2}}.
}

The sum in the square root is positive in some finite region in the space of running couplings containing the origin $\lambda = 0, \lambda_{y} = 0$---the free fixed point. In particular, it is positive at weak coupling. The $R$--charges are therefore always $R_i \leq 1$. It may happen that some field develops sufficiently large, and positive anomalous dimension along a flow (at strong coupling) such that its $R$--charge $R_i > 1$ \AmaritiWC. In this case it is argued in \AmaritiWC\ that one has to take the other positive branch of the square root in \rnmx. However, it is shown in \refs{\AmaritiWC\HookFP-\IntriligatorXXA} that even in this case it cannot surpass the upper bound $4/3$, therefore $R_i \le {4/ 3}$.

In the regime $R_i > {4/3}$, in general, we are in the strong coupling regime and the description in terms of the original infrared degrees of freedom often breaks down \refs{\KutasovIY}. A simple example in which a breakdown of degrees of freedom occurs is the electric SQCD theory with $N_f < 3N_c/2$ \refs{\KutasovIY}. Thus, in a regime in which the theory is internally consistent, it is hardly possible to meet the requirement that at least one of the $R$--charges should be larger than $5/3$. Our discussion, therefore, strongly suggests that asymptotic safety does not occur in $\N = 1$ supersymmetric gauge theories with or without superpotentials, and in particular, at weak coupling.

\bigskip\bigskip
\noindent{\bf Acknowledgements:}

I thank D. Kutasov for useful discussions and comments. This work is supported in part by DOE grant DE-SC0009924.

\appendix{A}{Dangerously irrelevant and harmlessly relevant operators}

Consider deforming an ultraviolet conformal field theory (an ultraviolet CFT) by adding to its Lagrangian a relevant operator. This results in an RG flow away from the ultraviolet fixed point. It may happen that as we flow toward low energies, an operator that was initially irrelevant (relevant)\foot{The operator can also be marginal.} at the ultraviolet fixed point builds up negative (positive) anomalous dimension, and eventually becomes relevant (irrelevant) in the infrared. Such an operator that becomes relevant (irrelevant) in the infrared is sometimes referred to as a dangerously irrelevant (harmlessly relevant) operator. Below, we review a few examples of this RG phenomenon in $\N = 1$ supersymmetric gauge theories in four dimensions.

Consider $\N = 1$ supersymmetric QCD (SQCD); an $SU(N_c)$ gauge theory with $N_f$ flavors of chiral superfields $Q^i$ and $ \tilde{Q}_{\tilde{i}}$ $(i, \tilde{i} = 1, \cdots, N_f)$ in the fundamental and anti--fundamental representations of $SU(N_c)$ respectively. For $N_f < 3N_c$, the theory is asymptotically free in the ultraviolet, \ie\ it describes free quarks and gluons, and it flows in the infrared into a non--trivial fixed point. Now consider adding the superpotential $W = h\left(Q\tilde{Q}\right)^2$. At zero gauge coupling $g$, the superpotential coupling $h$ is irrelevant and therefore it flows to zero in the infrared. On the other hand, if we turn on the gauge coupling $g$, and if $N_f < 2N_c$, then $h$ becomes relevant as we flow toward low energies \refs{ \StrasslerQS}. Thus, in this regime of $N_f$ the coupling $h$ is a dangerously irrelevant coupling.

Another example of dangerously irrelevant coupling is the gauge coupling $g$ in supersymmetric QCD (SQCD) with $N_f > 3N_c$. In this regime of $N_f$, the gauge coupling $g$ is marginally irrelevant; therefore, the theory is free in the infrared. Now consider giving masses to some of the quarks by adding the superpotential $W = \sum_i^nm_iQ_i^2$. In the large mass limit, the heavy quarks decouple from the dynamics. This reduces the number of fundamental quark flavors from $N_f$ to $N_f - n$. If the number of light quarks $N_f  - n < 3N_c $, then the gauge coupling $g$ becomes marginally relevant. Therefore, it is a dangerously irrelevant coupling in this limit.

A further example of the phenomenon appears in the Seiberg dual description of $\N = 1$ $SU(N_c)$  SQCD. The dual description has gauge group $SU(N_f - N_c)$, $N_f$ flavors of chiral superfields $q_i$ and $\tilde{q}^{\tilde{i}}$ $(i, \tilde{i} = 1, \cdots, N_f)$ in the fundamental and anti--fundamental representations of the gauge group respectively, and $N^2_f$ singlet ``meson'' chiral superfields $M_{\tilde{i}}^{j}$ $(j, \tilde{i} = 1, \cdots, N_f)$. The meson fields couple to the fundamentals $q$, $\tilde{q}$ through the superpotential $W = yMq\tilde{q}$. In the conformal window, \ie\ for $N_f$ in the range $3N_c/2 < N_f < 3N_c$, both the original and dual theories are asymptotically free in the ultraviolet, and in the infrared they both flow into non--trivial fixed points which are identified by the duality \refs{\SeibergPQ}. At zero gauge coupling $g$, the Yukawa coupling $y$ is marginally irrelevant and therefore it switches off as we flow toward low energies. However, if we turn on gauge interactions, $y$ becomes relevant as we flow toward the infrared fixed point \StrasslerQS. Thus, the Yukawa coupling $y$ is a dangerously irrelevant coupling. An example of a harmlessly relevant coupling is $t$ in the meson mass matrix $W = tM^2$, which is the magnetic dual of the superpotential $W = h\left(Q\tilde{Q}\right)^2$ considered earlier. At the free fixed point, that is in the ultraviolet, the mass matrix coupling $t$ is relevant. However if we turn on the (magnetic) gauge coupling $g$ and Yukawa coupling $y$, and assuming that  $N_f > 2N_c$, it becomes irrelevant in the infrared \StrasslerQS. Therefore, in this regime of $N_f$ it is a harmlessly relevant coupling.

Yet another example of the phenomenon arises in adjoint $\N = 1$ SQCD (ASQCD), \ie\ $\N = 1$ SQCD with an additional chiral superfield $X$ in the adjoint representation of the gauge group $SU(N_c)$. ASQCD is asymptotically free in the ultraviolet for $N_f < 2N_c$. Without a superpotential, turning on the gauge coupling $g$ drives the theory to flow into a non--trivial fixed point in the infrared \refs{\IntriligatorSM,\ \IntriligatorJJ}. Consider now the operator $\Tr X^{k + 1}$.  At the ultraviolet fixed point, that is at $g = 0$, it has, for $k > 2$, dimension larger than three, and hence it is marginal and/or irrelevant. However if we turn on $g$, for small $N_f$, it becomes relevant as we flow toward the infrared limit of the adjoint theory with $W = 0$ \refs{\KutasovNP, \ \KutasovSS}. Thence, in this regime of $N_f$, it is an example of a dangerously irrelevant operator.

The adjoint SQCD considered above also admits harmlessly relevant operators  \refs{\KutasovUX, \ \KutasovIY,  \ \IntriligatorJJ}. Consider ASQCD with $W = 0$ at its infrared fixed point in a regime where the operator $\Tr X^{k + 1}$ is relevant. We are now interested in deforming the theory by adding the superpotential $W = g_{k}\Tr X^{k + 1} + g_{k'} \Tr X^{k' + 1}$. Consider first turning on only the coupling $g_k$. The operator $\Tr X^{k + 1}$ will then drive the theory into a fixed point which we will call $F_k$. At this fixed point, the $R$--charge of the superfield $X$ is $R_X = 2/\left(k + 1\right)$. Therefore with $k' < k$ the operator $\Tr X^{k' + 1}$ is relevant at $F_k$. Now if we turn on the other coupling $g_{k'}$, then the theory at $F_k$ will evolve further into a new fixed point $F_{k'}$ at which the effective superpotential is $W \approx g_{k'}\Tr X^{k' + 1}$. At $F_{k'}$ the operator $\Tr X^{k + 1}$ is therefore irrelevant, and thus it is a harmlessly relevant operator \refs{\KutasovUX, \ \KutasovIY, \ \IntriligatorJJ}.

\appendix{B}{$a$--function and RG flows}

In this appendix we briefly review the generalized central charge $a$--function $a(\lambda_i)$ that is constructed in \KutasovUX\ (see also \BarnesJJ) by extending the method of $a$--maximization \IntriligatorJJ\ away from fixed points of an RG flow. It is an analogue of the Zamolodchikov $c$--function in four dimensions, and it manifestly satisfies the weak $a$--theorem, $a_{\rm UV} > a_{\rm IR}$. 

In the following, we first outline the procedure for determining the form of the generalized central charge $a$--function $a(\lambda_i)$ of a four dimensional $\N = 1$ supersymmetric gauge theory. We then show how the generalized central charge $a$--function $a(\lambda_i)$  can be used to analyze an RG flow between an ultraviolet, and an infrared fixed points of the theory. 

We consider a supersymmetric gauge theory with a gauge group $G$ and coupling $g$ containing chiral superfields $\Phi_i$ in irreducible representations $r_i$. We take a superpotential of the form 
\eqn\sup{W = y\prod_i\left(\Phi_i\right)^{n_i}.
}
 We write more conveniently the gauge coupling $g$ as $\alpha := \frac{g^2}{4\pi} $, and the coupling $y$ as $\alpha_{y} := \frac{y^2}{4\pi}$. There are two steps in finding the general form of the generalized central charge $a$--function $a(\lambda, \lambda_{y})$ of the theory along an RG flow.

The first step is constructing the following intermediate $a$--function; by abuse of notation we use the same letter $a$, 
\eqn\at{\eqalign{a(\lambda, \lambda_{y}, R_m)=\ &2|G| + \sum_i |r_i|\left[3(R_i - 1)^3 - (R_i - 1)\right]\cr
- \ &\lambda\left[T\left(G\right) + \sum_i T(r_i)\left(R_i - 1\right)\right]\cr
-\ &\lambda_{y}\left(2 - \sum_in_iR_i\right),
}}
 where $|r_i|$ and $T(r_i)$ are the dimension and index of the irreducible representation $r_i$ respectively. $|G|$ is the dimension of the gauge group $G$, and $T(G)$ is the index for the adjoint representation of the gauge group $G$. For example, for $G = SU(N)$,  $T({\bf{N}}) = T({\bar{\bf{N}}}) = 1/2$, $T(G) = N$. The Lagrange  multiplier $\lambda$ imposes the constraint that the superconformal $U(1)_R$ symmetry (at the non--trivial fixed points) is anomaly free (see also \ErkalSH\ for a review); $\lambda_{y}$ is the Lagrange  multiplier for the constraint that the general $R$ symmetry is preserved by the superpotential $W$.\foot{This can be easily generalized for theories based on product gauge groups, and consist of more that one superpotential coupling by introducing as many Lagrange multipliers as there are couplings.}
 
The second step is finding the unknowns $R_i$ and substituting them back into \at.  The $R_i$ are obtained by extremizing the intermediate $a$--function \at\ with respect to $R_i$ \IntriligatorJJ, which gives
\eqn\rnmx{R_i(\lambda, \lambda_{y}) = 1 - \frac{1}{3}\left(1 + \frac{\lambda T(r_i) - \lambda_{y}n_{i}}{|r_i|}\right)^{\frac{1}{2}}.
}
Substituting the $R_i$ \rnmx\ back into \at\ yields the generalized central charge $a$--function $a(\lambda, \lambda_{y})  := a(\lambda, \lambda_{y}, R_i(\lambda, \lambda_{y}))$ of the theory. Because $R_i(\lambda, \lambda_{y})$ solves $\partial a/\partial R_i = 0$, one finds that
\eqn\lbeta{\eqalign{\frac{da}{d\lambda} \ & =  - \left[T\left(G\right) + T(r_i)\left(R_i - 1\right)\right] := \hat{\beta}_{G}(\lambda, \lambda_y), \cr
\frac{da}{d\lambda_{y}} \ & =  -\left(2 - \sum_in_iR_i\right) := \hat{\beta}_y(\lambda, \lambda_{y}),
}}
here the $R_i$ are given by \rnmx. 

Since $\lambda, \lambda_{y}$ were originally introduced as Lagrange multipliers, we must extremize the generalized  central charge $a$--function $a(\lambda, \lambda_{y})$ with respect to the multipliers $\lambda, \lambda_{y}$. We note that there are two cases to consider. We first consider the case in which both the couplings are turned on; and next, that only one of the couplings is turned on. 

In the former case, we must extremize $a(\lambda, \lambda_{y})$ with respect to both $\lambda, \lambda_{y}$. This gives 
\eqn\lbetaz{\hat{\beta}_G(\lambda, \lambda_y) = 0, \qquad \hat{\beta}_y(\lambda, \lambda_{y}) = 0.
}
In the latter case, if (say) only $\alpha$ is turned on, then $\lambda_y = 0$ and extremizing $a(\lambda, 0)$ with respect to $\lambda$ gives
\eqn\lbetazz{\hat{\beta}_G(\lambda, 0) = 0.
}
On the other hand, if only $\alpha_y$ is turned on, then $\lambda = 0$ and extremizing $a(0, \lambda_y)$ with respect to $\lambda_y$ gives
\eqn\lbetazzz{\hat{\beta}_y(0, \lambda_y) = 0.
}

These equations fix the multipliers $\lambda, \lambda_y$ and in turn also $R_i(\lambda, \lambda_{y})$ and $a(\lambda, \lambda_{y})$. We will denote the values of the multipliers which solves \lbetaz, \lbetazz, \lbetazzz\ by the same letters $\lambda^*, \ \lambda_y^*$ in all these different cases; this should cause no ambiguity. We assume that $\lambda^*, \ \lambda_y^*$ are such that the beta function for the coupling $\alpha$ is not singular in the regime $\alpha \leq \alpha^*$; the construction of \KutasovUX\ is valid only in a certain finite region in the space of couplings containing the origin. 

The $R_i$ \rnmx\ at the fixed points $\lambda = 0, \lambda_y = 0$ and $\lambda = \lambda^*, \lambda_{y} = \lambda_y^*$ give the $R$--charges of the chiral superfields $\Phi_i$ associated with the $U(1)_R$ superconformal symmetry. Similarly, the generalized central charge $a(\lambda, \lambda_{y})$ at the fixed points gives the central charges $a_{UV}$, $a_{IR}$ with $a_{UV} > a_{IR}$.

The Lagrange multipliers $\lambda, \lambda_{y} $ are identified in the construction of \KutasovUX\  in some renormalization scheme with the running couplings $\alpha, \alpha_{y}$; therefore, they are always positive.  In the vicinity of $\alpha = 0$, $\alpha_{y} = 0$ it was shown in \refs{\KutasovUX\BarnesJJ-\KutasovXU} that  $\lambda \approx |G|\alpha + \O(\alpha^2)$, $\lambda_{y} \approx \alpha_{y} + \O(\alpha^2)$.  The positivity of the Lagrange multipliers gives non--trivial constraints on asymptotic safety. 
 
 The identification of $\lambda, \lambda_{y}$ with the running couplings $\alpha, \alpha_{y}$ suggests the interpretation of $R_i(\lambda, \lambda_{y})$ and $a(\lambda, \lambda_{y})$, in some renormalization scheme, as running $R$--charges and central charge. Thus, in \lbeta\ we note that, in the vicinity of $\lambda = 0$, $\lambda_{y} = 0$, $\hat{\beta}_G(\lambda, \lambda_y)$ is proportional to the beta function for the coupling $\lambda$, and $\hat{\beta}_y(\lambda, \lambda_{y})$ is proportional to the beta function for the coupling $\lambda_{y}$.\foot{In general one has $\frac{da}{d\lambda^i} = G_{ij}\beta^{j}$, where $G_{ij}$ is the metric on the space of couplings, and $\beta^i$ are the beta functions.} We also note that this interpretation generalizes the method of $a$--maximization \IntriligatorJJ\ away from fixed points of an RG flow.

The equations \lbeta, together with the positivity constraint, determine RG fixed points in the space of running couplings $\lambda, \lambda_{y}$ (in some renormalization scheme). In some finite domain in the space of couplings containing $\lambda = 0, \lambda_{y} = 0$, the fixed points of RG flows are found by setting \lbeta\ to zero, \ie\ by solving the corresponding equation(s) among \lbetaz, \lbetazz, \lbetazzz. 

We note that the gauge coupling $\lambda$ is relevant throughout an RG flow (away from the free fixed point) if $\hat\beta_G(\lambda, \lambda_y) < 0$. Therefore, if $\lambda$ is relevant at the fixed point \lbetazzz, then turning it on will drive the theory further down in the infrared into the fixed point \lbetaz\ provided $\lambda_y$ is not harmlessly relevant. In the case in which $\lambda_y$ is harmlessly relevant we flow instead into the fixed point \lbetazz. 

We also note that away from the free fixed point the coupling $\lambda_{y}$ is relevant along an RG flow if $\hat\beta_y(\lambda, \lambda_y) < 0$. Similarly, if $\hat\beta_y(\lambda, \lambda_y) < 0$, then one can turn on $\lambda_y$ at the fixed point \lbetazz, and flow further down in the infrared into the fixed point \lbetaz. if, however, $\lambda$ is harmlessly relevant, then we flow instead into the fixed point \lbetazzz. 

Turning on both couplings at the free fixed point in the case in which they are (marginally) relevant (and dangerously irrelevant) leads in general in the infrared into the fixed point \lbetaz. If either of the couplings is harmlessly relevant, then it leads into the fixed point \lbetazz\ or \lbetazzz\ (see the preceding two paragraphs).

As a simple example to illustrate the procedure, we consider a class of $\N = 1$ supersymmetric gauge theories with a gauge group $SU(N_c)$ coupled to $N_f$ (anti--)fundamental chiral superfields ($\widetilde{Q}$)$Q$, and a bi--fundamental gauge singlet chiral superfield $M$. It has $SU(N_f)_L\times SU(N_f)_R\times U(1)_R$ global symmetry. We summarize the field contents, and the representations under which they transform as follows.

 $$
\vbox{\tabskip=0pt
\halign{\tabskip=0.2cm #\hfil&#\hfil&#\hfil&#\hfil&#\hfil&#\hfil&#\hfil&#\hfil&#\tabskip=0pt \cr
Chiral superfields & $SU(N_c)$&$SU(N_f)_L$&$SU(N_f)_R$\cr
$Q_i(\widetilde{Q}_{\widetilde{i}})(i,{\widetilde{i}} = 1, \cdots, N_f)$ &$\bf N_c(\bar N_c)$&$\bf 1(\bf \bar{N}_f)$&$\bf N_f(\bf 1)$\cr
$M_{\widetilde{i}i}({\widetilde{i}}, i = 1, \cdots, N_f)$&$\bf 1$&$\bf N_f$&$\bf \bar{N}_f$\cr}}
$$

We take the superpotential
\eqn\supyu{W = y\Tr \widetilde{Q} M Q.
}

The first step in finding the $a$--function, and its gradient flow equations is constructing the intermediate $a$--function \at. In this class of theories, it follows from \at, the intermediate $a$--function takes the form
\eqn\afuncT{\eqalign{a(\lambda,R_j)=\ &2\left(N_c^2-1\right)\cr
+ \ &\ 2N_f\cdot N_c\cdot \left[3(R_Q-1)^3-(R_Q-1)\right]
+ \ N_f^2\cdot 1\cdot \left[3(R_M-1)^3-(R_M-1)\right]\cr
- \ &\lambda\left[N_c+N_f(R_Q-1)\right]\cr
- \ &\lambda_y\left(2-2R_Q-R_M\right),
}}
where $R_{\widetilde{Q}} = R_Q$; the $U(1)_R$ symmetry commutes with the chiral symmetry $SU(N_f)_L\times SU(N_f)_R$.

The next step is extremizing \afuncT\ with respect to the unknowns $R_i$. This gives a set of equations that can be solved for $R_i$ in terms of the couplings $\lambda, \lambda_y$. We obtain the running $R$--charges $R_i(\lambda, \lambda_y)$
\eqn\rrex{\eqalign{
R_Q(\lambda, \lambda_y) = \ &1-{1\over3}\left(1+{\lambda N_f-2\lambda_y\over 2N_fN_c}\right)^{\half},\cr
R_M(\lambda, \lambda_y) = \ &1-{1\over3}\left(1-{\lambda_y\over N_f^2}\right)^{\half}.
}}

Substituting these into \afuncT\ finally gives the generalized $a$--function $a(\lambda, \lambda_y)$. The gradient flow equations of the $a$--function are then obtained using \rrex\ in \lbeta. For small couplings $\lambda_1, \lambda_y$, for instance, these flow equations to leading order in the couplings are
\eqn\lbetaT{\eqalign{\hat{\beta}_{G}(\lambda, \lambda_y) \ & = -{N_c\over 3}\left[(3 - x)  - {1\over 4} x {\lambda \over N_c} + {1\over 2}x^2{\lambda_y\over N_f^2}\right] , \cr
\hat{\beta}_y(\lambda, \lambda_{y}) \ & =  {1\over 6}\left[\left(2 {x} + 1\right) {\lambda_y\over N_f^2} - {\lambda\over N_c}\right],
}}
here $x := {N_f \over N_c}$. With $\alpha_g := \frac{N_cg^2}{(4\pi)^2}, \alpha_y := \frac{N_cy^2}{(4\pi)^2}$, to leading order in $\alpha_g, \alpha_y$ we have $\lambda = 8N_c\alpha_g, \lambda_y = 4N_f^2\alpha_y$ \refs{\KutasovUX\BarnesJJ-\KutasovXU}.
\listrefs
\end